\documentclass[preprint,review,times,3p]{elsarticle}
\usepackage{graphicx}
\usepackage{subfig}
\usepackage[countmax]{subfloat}
\usepackage{amsmath, amssymb}

\biboptions{sort&compress,comma}

\journal{ Physica E}

\begin{document}

\begin{frontmatter}

\title{Single-electron Transport Through Quantum Point Contact}

\author[label1]{G. Bilge\c{c} Aky\"uz\corref{label4}}
\ead{gonulbilgec@adu.edu.tr} \cortext[label4]{Corresponding Author.
Tel.: +902562128498; Fax: +902562135379} 
\author[label2]{A. Siddiki}

\address[label1] {Adnan Menderes University, Faculty of Arts and
  Sciences, Physics Department, 09100, Ayd{\i}n, Turkey }
\address[label2]{Mimar Sinan Fine Arts University, Faculty of Science and Letters, Physics Department, Bomonti-Sisli, 34380 \.Istanbul,
Turkey}
\begin{abstract}
Here, we employ a numerical approach to investigate the transport
and conductance characteristics of a quantum point contact. A
quantum point contact is a narrow constriction of a width comparable
to the electron wavelength defined in a two-dimensional electron gas
(2DEG) by means of split-gate or etching technique. Their properties
have been widely investigated in the experiments. We define a
quantum Hall based split-gate quantum point contact with standard
gate geometry. Firstly, we obtain the spatial distribution of
incompressible strips (current channels) by applying a self
consistent Thomas-Fermi method to a realistic heterostructure under
quantized Hall conditions. Later, time-dependent Schrodinger
equation is solved for electrons injected in the current channels.
The transport characteristics and time-evolutions are analyzed in
the integer filling factor regime ($\nu=1$) with the single electron
density. The results confirm that the current direction in a
realistic quantum point contact can be controllable with the
external interventions.
\end{abstract}

\end{frontmatter}

\section{Introduction}
\label{intro}

Developments in the quantum information processing technology have
lead to intensive studies in the investigation of intrinsic
properties of small-scale electronic devices. The conductance
quantization is a fundamental phenomenon of electron transport in
the low-dimensional structures and can be observed in a quantum
point contact (QPC) which is constructed by geometric or
electrostatic confinement of a two-dimensional electron gas.
Therefore, a detailed understanding of the QPC conductance which
represents cornerstone of mesoscopic physics and is of the prime
importance via conductance through them is quantized at zero
magnetic field \cite{Siddiki2007}. So the properties of these
small-scale devices have been generously investigated in the
experiments such as Quantum Hall effect of the Mach Zender
\cite{Ji03,Neder06}, Aharanov-Bohm interferometers and the
observation of $0.7$ anomaly \cite{Thomas1996_1,Thomas1996_2}.
Electrostatic and transport properties of the QPC's in quantized and
non-quantized magnetic fields are also investigated early studies,
comprehensively \cite{Siddiki2007}. In a Quantum Hall device, the
current carrying states are compose of with the Landau Level
quantization and electron distribution in a Hall device depends on
the pinning of the Fermi level to the highly degenerate Landau
levels. According to self-consistent screening theory, 2DES contains
two different kinds of regions namely incompressible (IS) and
compressible strips (CS) \cite{Chklovskii92, Siddiki2003}, which
means if the Fermi level lies within a Landau level with the high
density of state, system is called as compressible otherwise system
is called as incompressible. In the compressible region, system has
a flat potential profile and the screening property. In case of
incompressible strips, system has constant electron density and
generally a spatially varying potential because of the absence of
the screening \cite{Siddiki2007}. The current is transported by the
scattering-free ISs when the quantum mechanical effects are smaller
than the ISs widths \cite{siddiki2004}. Therefore, it is very
important to deal with realistic constraining geometries due to
boundary condition dictated properties of incompressible strips.

In the present study, we aimed to obtain a comprehensive
understanding of the conductance of a standard QPC geometry in a
quantum Hall device. We solved the 3D Poisson equation
self-consistently and utilized Thomas-Fermi approximation to a
heterostructure \cite{arslan} taking into account the
lithographically defined surface gates and we obtained the electron
and potential distributions and incompressible strips
($\nu=1$-current channels) under quantum Hall conditions. To
describe the current channels, we use a model potential and a
time-dependent solution to observe the conductivity of a wave packet
injected into the channel. We have searched how is affected the QPC
conductance with the interchannel distance, magnetic flux, channel
depth and channel width. Additionally, we investigate and seek for
the optimal conditions of a single-electron transport in a QPC
geometry.

\section{Model and Methods}
\label{model}
\begin{figure}[ht!]
\centering
\includegraphics[width=0.8\columnwidth]{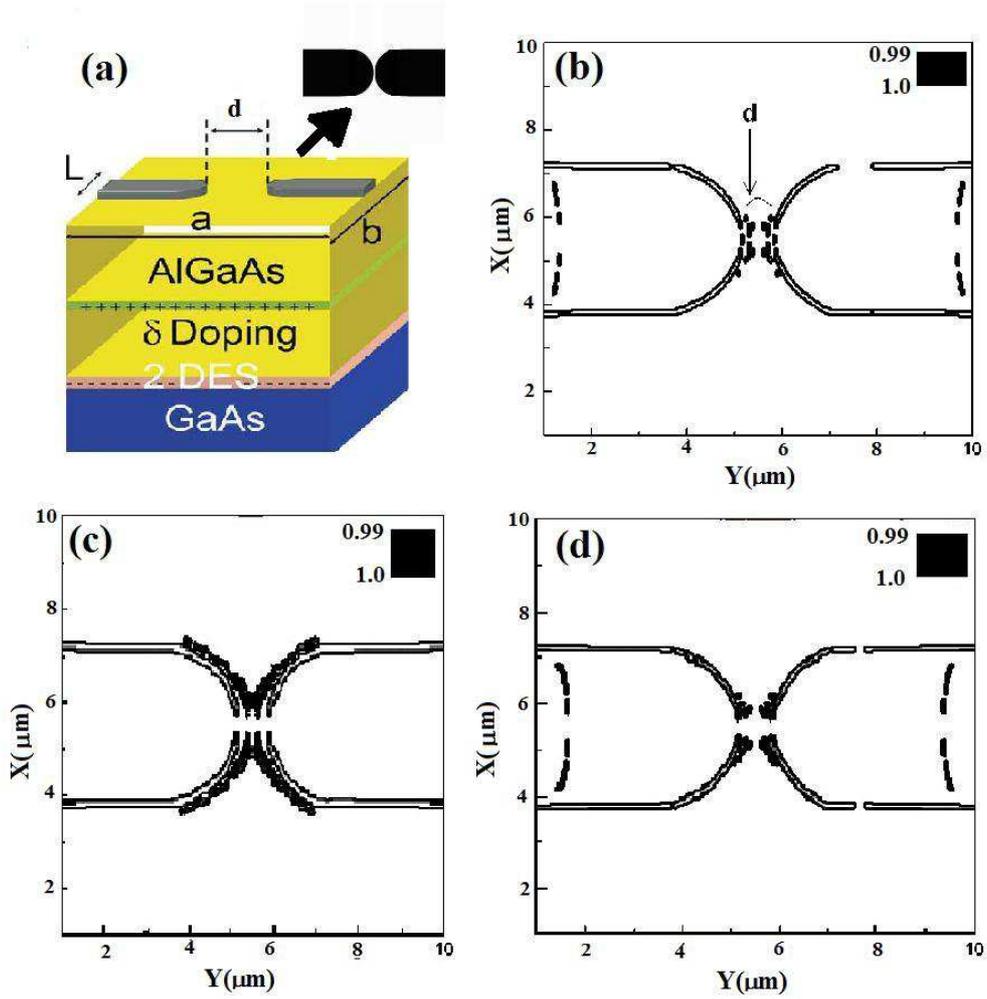}
\caption{Spatial distribution of the incompressible strips
calculated at (a)Heterostructure, the metallic gates are deposited
on the surface. (b)Vgate=1.7, B=2.5T, (c) Vgate=1.7, B=5 T (d)
Vgate=2.6, B=2.6 T. It is expected that only one can observe the
conductance at (d) conditions, whereas, the scattering between the
edge states is presented in other cases.} \label{fig:ISs}
\end{figure}

The structural information is provided by Arslan et al \cite{arslan}
which consists of metallic surface gates (dark semielliptic regions)
and a thin donor layer ($\delta$-silicon doping) which provides
electrons to the two-dimensional electron gas (2DES)
(Fig.\ref{fig:ISs}a). 2DES is confined to a thin area which is
located at the interface of the heterostructure. The spatial
distribution of the electron density is determined by the donors,
the gate voltage $V_g$ and the gate shape. In a Quantum Hall device,
the current carrying states result from the Landau level
quantization, followed by level bonding at the edges. So the
transport takes place through the edge states. The potential and
charge distribution of the system under quantum hall conditions can
be obtained by solving poisson equation self-consistently within
Thomas-Fermi approximation. Electrostatic calculations are performed
as in refs \cite{Kotimaki-engin,arslan} and the transport simulation
has been carried by injecting single-electron wave packet to a
static 2D potential.

In Figure~\ref{fig:ISs}/b-d, we show the spatial distribution of the
incompressible strips (ISs) at different magnetic fields and gate
potentials. As explicitly seen, magnetic field
(Fig.\ref{fig:ISs}/b-c) and gate potential ($V_g$) determine the
spatial distribution of the ISs (Fig.\ref{fig:ISs}/d). The
interchannel distance decrease while the ISs width increases with
the increasing magnetic field. At lower magnetic fields, the strips
are far apart and the conductance are weak or not present. So, it is
expected that the wave packet remains in the incoming channel (lower
left channel). To analyze the effects of internal distance, channel
depth and channel width, we focus on realistic modeling of current
channels along ISs following by a dynamical study on electron
transport in the device. The channels are modeled by a 2D potential
profile consisting of two curved pipes following the shape of the
ISs in Figure (\ref{fig:ISs}/b-d). The distance between the left and
right channels is varied as $d=0-1$ range (in atomic units).

The potential to transport corresponds to (ISs) with $\nu=1$ integer
filling factor and these ISs carry the unbalanced electrical
current. The potential minima of the channels are shown as solid
lines in Figure \ref{fig:ISs}/b-d and the cross section of these
channels have a Gaussian form;
\begin{equation}
V=-V_0exp(-s^2/c^2),
\end{equation}
where s is the perpendicular axis to the current-channel. $V_0=20$
(Hartree a.u. used throught) is the depth of the track which prevent
the electrons from escaping the channels and c is the variable width
parameter, initially set to $0.2$. It is useful to choose a gaussian
form for the current channels because, it is a reasonable
approximation for the magnetic enclosure at the bottom of the
channel and this form enable to leak electron current coherent with
the experimental case. Additionally, in the time-propagation, excess
electron density on top of the Fermi background is used and is
assumed to be static.


In the calculations, as an initial condition we set a
single-electron wave packet in the lower-left corner of the
incompressible strip and we track the density during the time
propagation until we find back scattering from the corners of our
finite simulation box. Thus, the direction of the current is the
same in right and left channels (non-chiral transport) as in the
non-equilibrium transport experiment. To provide a source-drain
voltage to the current channel, it is applied a linear ramp
potential ($V=-0.2V_0$) which accelerate the initial state. In the
transport experiments, an external current is applied and a Hall
potential is observed and this potential has the same slope in the
two opposite corners. Therefore a potential drop occurs at the
incompressible strips which are also observed in the some
experiments \cite{Ahlswede01, Ahlswede02}.

The conductance as a function of the magnetic flux, the channel
depth, interchannel distance and channel width is determined by
calculating the probability $N_r$ to find the electron in the
corners within the determined simulation time ($t=14$ a.u)
\cite{Kotimaki}. Time dependent density functional theory (TDDFT) is
applied  upon a 2D-real space grid for the time propagation of the
electrons. The calculations have been performed by the OCTOPUS code
package \cite{MAL_Marques}. In the calculations, independent
electron approximation has been used to completely neglect the
interactions by using single electron wave-packet.

\section{Results}
\label{results} Transport simulation is carried out by injecting
single-electron wave packet into a static 2D potential. The
potential minima of the channels (Fig.\ref{fig:ISs}/b-d) are modeled
by a 2D potential profile following the shape of the IS. We consider
that the interchannel distance (d) would be essentially controlled
either with gates or magnetic field.

\begin{figure}[h!]
\centering
\includegraphics[width=1.0\columnwidth]{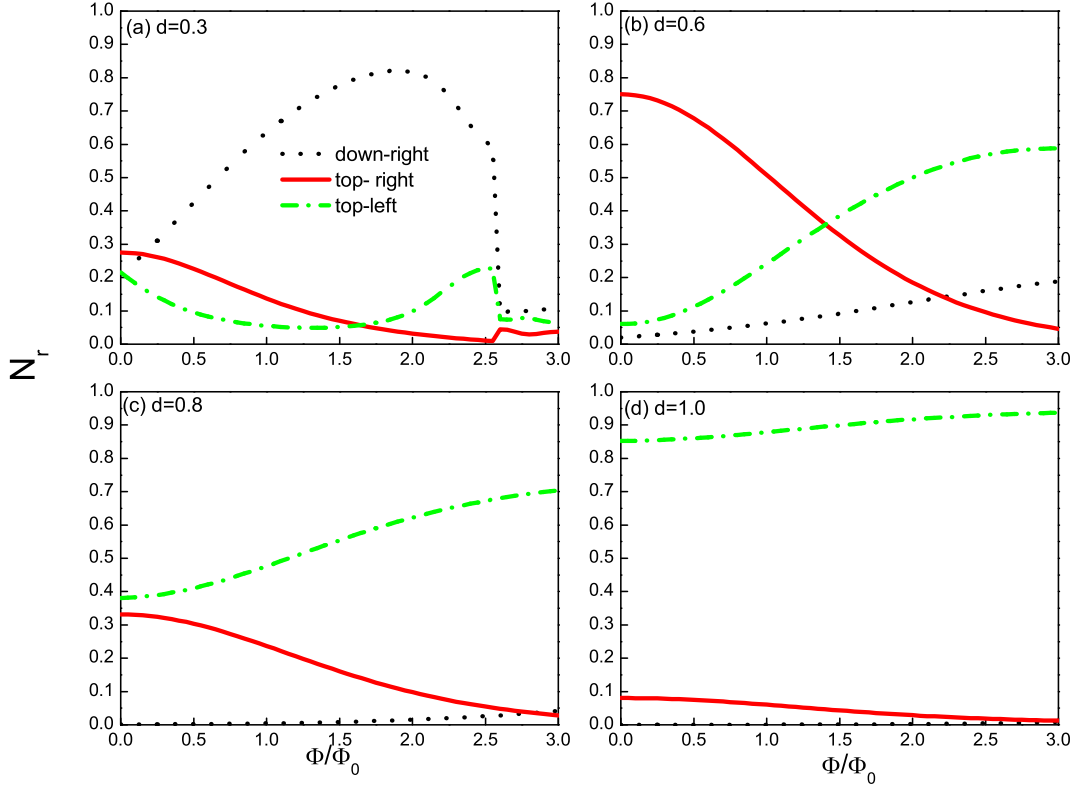}
\caption{Estimated conductance ($N_r$) as a function of the magnetic
flux in the center of QPC-gates for different interchannel
distances.} \label{fig:fig2}
\end{figure}

In our transport simulations, we track electron dynamics as a
function of a magnetic flux which is applied the center of
interchannel distance and added to the background magnetic field. In
the first time-dependent simulations, we applied relative fluxes
$\phi/\phi_0$ which is used in the range of $0.0-3.0$. In here,
$\phi_0=h/e$ and $\phi=B\times A$, A define an area in which
magnetic field is applied. The conductance of the system is
considered as a function of the uniform and perpendicular magnetic
field strength in the units of the enclosed flux quanta
($\phi/\phi_0$). The wave packet with slope $V_1=-0.2 V_0$ is
accelerated along the current channels. To analyze the transport
characteristics in detail, we calculate the conductance as a
function of the magnetic flux for different interchannel distances
($d=0.3-1.0$) in Figure\ref{fig:fig2}.

The panels in Figure \ref{fig:fig2} show conductance for different
interchannel distances at the corners. In this analysis, we compare
the electron density at the different corners (top-right, down and
top-left) and obtain distinctive characteristics for different
conditions. Maximum probability has been observed in the lower-right
channel conductance for the minimum interchannel distance (d$=0.3$)
(Fig.2a). It is observed a small peak at nearly $\phi/\phi_0=2.6$
value for the top-right conductance and at this critical magnetic
flux value, conductance probabilities of other current channels show
a sharp decline. Magnetic flux applied the center of the
interchannel distance partly prevent to transport to the right
channel  and go ahead to top-left above $\phi/\phi_0=2.6$ value.
Maximum conductance is obtained at nearly $\phi/\phi_0=2$ value, for
down-right channel. A similar drop is also observed for the top-left
channel conductance for this interchannel distance. This value can
be called as critical current value because it is observed a
considerable decline in the probabilities of the channels above the
critical value. After this critical current value, largely amount of
single-electron wave packet remains in the lower-left channel
(incoming channel), but nevertheless probability do not vanish in
the three region. Conductance characteristic of the QPC with the
$d=0.6$ interchannel distance are more intelligible than $d=0.3$
(Fig.2(a)-(b)). In this situation, top-right conductance is
decreased while the left-top one is increased with the increasing
magnetic flux which is applied the center of the interchannel
distance. This magnetic flux prevent the transport of the electron
wave-packet to the top-right channel. Although, the transport schema
of the $d=0.8$ has similar characteristics with the transport of the
$d=0.6$, a discrepancy in the conductance graph is also observed for
$d=0.8$ interchannel distance. For $d=0.8$, conductance
probabilities are lower than the $d=0.6$ interchannel distance. The
changing rates of the right and left top channel conductance
probabilities are almost seen equal. It is also seen that the
probability ($N_r$) of the finding the electron in the right-down
corner is very weak.

For $d=1.0$ interchannel distance (Fig.\ref{fig:fig2}d), it is not
observed a important effect of the magnetic flux on the transport to
the right channels. Single-electron wave packet remain the left
channel and probability which has already small value decrease with
the increasing magnetic flux. So, in these conditions, interchannel
transport is completely blocked near the $\phi/\phi_0\approx3.0$
value. This behavior is explained by the reason of wide interchannel
distance ($d=1.0$) for the electron wave packet to the right
channels.


\begin{figure}[h!]
\centering
\includegraphics[width=0.5\columnwidth]{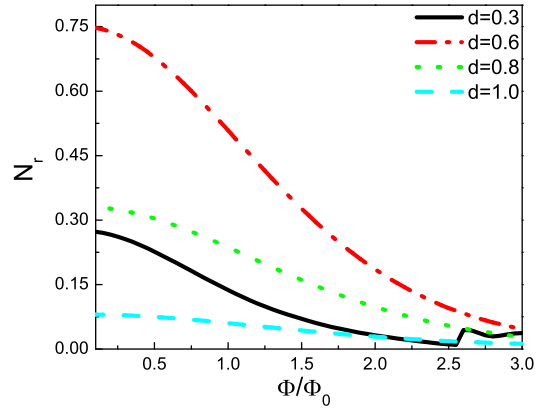}
\caption{Estimated conductance (top-right channel) ($N_r$)-flux
graph at different distances} \label{fig:Nr_right_up_c}
\end{figure}

To determine the optimal interchannel distance for conductance which
correspond here to the electron density transferred to the top-right
channel, we plot the conductance probability ($N_r$) as a function
of the magnetic flux for different interchannel distances (Figure
\ref{fig:Nr_right_up_c}). Optimal distance is determined as d$\sim
0.6$ interchannel distance from this analysis. At smaller and larger
values of interval distances, conductivity probability is
significantly decreased. This result confirm similar transport
studies \cite{Kotimaki-engin, Aysevil}. In general, the top-right
conductance probability of the QPC devices decrease linearly with
the increasing magnetic flux applied to the center of QPC. However,
it is observed a peak for the $d=0.3$ interchannel distance above
the nearly $\phi/\phi_0=2.6$ magnetic flux value. The reason of the
observation such a peak, it may be sharp decline in the conductivity
for other current channels for this interchannel distance.

\begin{figure}[h!]
\centering
\includegraphics[width=0.5\columnwidth]{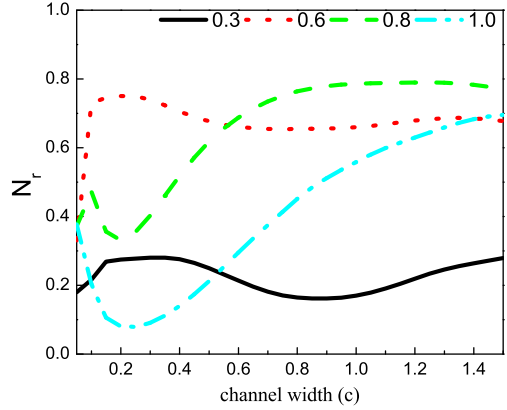}
\caption{Estimated conductance (top-right channel-$N_r$) at zero
magnetic field ($\phi/\phi_0=0$) as a function of channel width
($c$) at different interchannel distances (d), $V=0.2V_0$. The
optimal distance for the conductivity probability also depends on
channel width.} \label{fig:Nr_c}
\end{figure}

Conductance-channel width analysis show another property of the QPC
(Fig.\ref{fig:Nr_c}). It is seen that the optimal distance also
depends on the channel width. The conductivities of the devices with
the interchannel distances $d=0.6$ and $d=0.3$ show peaks while the
conductivities of the other devices show dips at nearly 0.2-0.3
channel width range. At the same time, probabilities vary smoothly
after the nearly $c=0.5$ value for $d=0.6$ and $d=0.8$.

Another remarkable observation of $N_r-c$ graph is that maximum
conductances are observed for d=$0.3-0.6$ interchannel distances
while minimum conductances are observed for d=$0.8-1.0$ interchannel
distances at small channel width ($\sim0.2$). Therefore, the optimal
top-right conductance probability is determined at nearly 0.2
channel width value for $d=0.6$ interchannel distance and at larger
channel widths ($c>0.5$) for $d=0.8$ interchannel distance. Above
the nearly $c=0.2$ value, conductance increase with the increasing
channel width in this $c$ range for $d=1.0$distance. Additionally,
it is seen that the conductance probabilities of $d=0.6$ and $d=1.0$
interchannel distance coincide for wide channel widths ($c>1.3$).
The reason of this coinciding is that the widening current channels
get closer to each other. Probability oscillate with small amplitude
with the increasing channel width for $d=0.3$ interchannel width.
Therefore, we emphasize that the channel width is as important as
interchannel distances on the QPC conductivities.

\begin{figure}[h!]
\centering
\includegraphics[width=0.5\columnwidth]{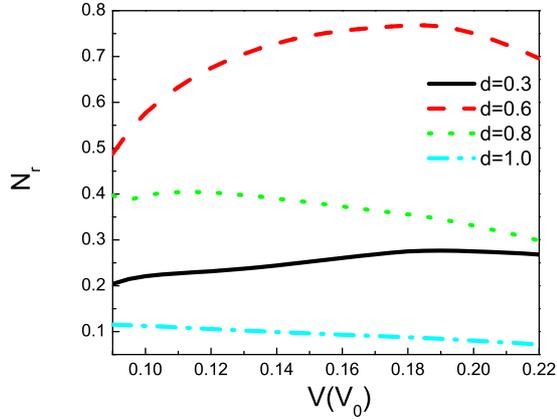}
\caption{Estimated conductance (top-right channel-$N_r$) at zero
magnetic flux ($\phi/\phi_0=0$) as a function of channel depth
values for c=0.2 values.} \label{fig:Figure5_alternatif}
\end{figure}

Finally, Figure \ref{fig:Figure5_alternatif} shows the top-right
conductance of the QPC for different channel depth values $V(V_{0})$
at zero magnetic flux value. The conductance probabilities increase
with increased potential depth for the devices with $d=0.3$ and
$d=0.6$ interchannel distances. Additionally, the results in Figure
\ref{fig:Figure5_alternatif} confirm that the optimal interval
distance is $0.6$ and maximum probability is obtained in $V=0.18V_0$
value for optimal interchannel distance. For $d=0.3$ interchannel
distance, changes in the conductance are non-crucial, $N_r$
oscillates with small amplitudes. Conductance probability to the
right channel decrease with the increasing channel depth for
$d=0.8$, $d=1.0$, because it is think that the single electron wave
packet are confined in the current channel. However, the conductance
probabilities almost linearly decrease with the increased channel
depth for the devices with $d=0.8$ and $d=1.0$ interchannel
distances. So, the increased channel depth has a negative effect on
the conductance for large interchannel distances. This confining of
electron wave packet is also affects the optimal interchannel
distance ($d=0.6$) for large channel depths and it is observed an
decreasing in the conductance probability. So, it can be concluded
that if the interchannel distances will be broad, channel depth must
be low for the high conductance probability.


\section{Conclusion}
\label{conclusion} We have performed a single-electron wave packet
propagation scheme  by using time dependent density functional
theory to study electron transport in the integer filling factor
($\nu=1$) regime for a realistic Quantum point contact geometry. We
observe distinctive QPC conductivity behaviors, where their
amplitudes strongly depend on the interchannel distances, channel
widths and channel depth values. For the top-right channel
conductivity, the interchannel distance has an optimal value
$d\sim0.6$  and $d\sim0.3$ for the down-right channel conductivity
when the conductance is maximized. However, we obtain an optimal
transport condition depending on channel width for the top-right
channel conductivity. Eventually, we obtain distinct conductivity
graphs in all cases and all directions. So, we conclude that the
direction and amplitude of a single electron wave packet is
completely controllable by the external factors as magnetic flux,
channel depth and width etc. We are sure that the numerical results
presented here will pave path for further experimental efforts in
the QPC conductivities.


\bibliographystyle{elsarticle-num}

\newpage







\end{document}